\documentclass[preprint]{revtex4}
\usepackage{graphicx}
\usepackage{hyperref}
\def\be{\begin{equation}}
\def\ee{\end{equation}}
\def\ba{\begin{eqnarray}}
\def\ea{\end{eqnarray}}

\def\ie{\textit{i.e. }}
\begin{document}
\parskip 3pt
\renewcommand{\topfraction}{0.8}
\vskip 3cm
\preprint{}
\title {\Large\bf Stringy corrections to a time-dependent background\\
 solution of string and M-Theory}

\author{\setcounter{footnote}{2}\vskip .7cm \bf Gustavo
  Niz$^{1,2,}$\footnote{\tt Gustavo.Niz@nottingham.ac.uk} and
Neil Turok$^{2,}$\footnote{\tt N.G.Turok@damtp.cam.ac.uk}} 
\affiliation{\vskip .5cm $^{1}${School of Physics and Astronomy, University of
  Nottingham, NG7 2RD, UK}\\ \\$^{2}${DAMTP, CMS, Wilberforce Road, 
  Cambridge, CB3 0WA, UK}\vskip .7cm} 
{\begin{abstract}  
\vskip .5cm We consider one of the simplest time-dependent backgrounds in 
M-theory, describing the shrinking away of the M-theory dimension
with the other spatial dimensions static. As the M-theory dimension 
becomes small, the situation becomes well-described by string theory
in a singular cosmological background where the string
coupling tends to zero but the $\alpha'$-corrections become large,
near the cosmic singularity.
We compute these $\alpha'$-corrections, both for the background and 
for linearized perturbations, in heterotic string theory, and show
they may be reproduced by a map from eleven-dimensional M-theory.
\end{abstract}}

\maketitle
\section{Introduction}
One of the most important tests for string and M-theory is to provide
a successful account of the cosmic singularity. While the singularity
almost certainly cannot be described in classical terms, there seem to
be two fundamentally different possibilities for its role in a quantum 
theory of gravity. Either the singularity marks the first emergence of
classical time, or it does not. In the former case, one must explain
why the classical universe emerged from the singularity in a dense, 
rapidly expanding state. In the latter case, one has to understand 
how a pre-big bang universe can propagate across the singularity 
and into a hot, expanding phase.

One of the simplest time-dependent backgrounds one can consider in M-theory
is compactified Milne spacetime, with metric
\be\label{metric11}
ds^2_{11}=-dt^2+t^2d\theta^2+\sum_{i=1}^9 (dx^i)^2, 
\ee
where $\theta$ parameterizes a circle of circumference $2\theta_0$.
In heterotic M-theory, which is of special interest for the ekpyrotic
and cyclic universe models \cite{ekp,cyclic} and shall be our main
focus here, the circle is modded out by $Z_2$, so that $0\leq \theta
\leq \theta_0$. 

For all $t\neq 0$, the metric is non-degenerate and the 
spacetime (\ref{metric11}) is flat, {\it i.e.} the Riemann 
curvature vanishes. Also, the 
four form field strength is zero. Therefore, 
if M-theory is describable in terms of local covariant field
equations for the background bosonic fields then (\ref{metric11})
automatically solves those equations.  

As suggested in Refs.~\cite{Seiberg} and \cite{turok}, analyticity in
$t$ suggests the possibility of extending  
the spacetime to the entire range $-\infty <t <\infty$ so that 
(\ref{metric11}) describes the collapse and 
re-expansion of the M-theory dimension, or in heterotic 
M-theory, the collision and separation of two orbifold planes. 
The conjecture that such a transition through $t=0$ is possible 
underlies the cyclic and ekpyrotic universe models \cite{cyclic,ekp}. 

When $t$ is small, the M-theory dimension is small and one should be 
able to describe the situation in terms of weakly coupled string theory.
Under the map from eleven dimensional M-theory to ten-dimensional 
string theory, at lowest order in $\alpha'$, (\ref{metric11}) 
yields a cosmological solution in the string frame,
\be\label{metrics}
ds^2_{string}=|t|(-dt^2+\sum_{i=1}^9 (dx^i)^2), \qquad\qquad
g_s=e^\phi=|t|^{3\over 2}, 
\ee
where $g_s$ is the string coupling and $\phi$ is the dilaton. 
Near $t=0$ stringy interactions are small and string loop 
effects may be neglected. However,
the string frame metric becomes singular so higher 
order corrections in $\alpha'$ become increasingly significant.  Also, 
as noted in \cite{Seiberg}, the background (\ref{metric11}) is 
not supersymmetric, hence there is no protection against large quantum
effects.  

The fact that the $\alpha'$ expansion fails certainly 
does not in itself 
mean that the background is not a good background for the 
quantization of strings.
In \cite{turok} it was shown that $M2$-branes wrapped across the
M-theory dimension, which reduce to fundamental strings as $t$ tends
to zero, generically  
obey regular classical evolution through $t=0$. It was 
argued that in the 
small $t$ regime the theory should be described by an expansion 
in $1/\alpha'$ ({\it i.e.} in the string tension), which unfortunately 
is not yet well understood. In \cite{niz} this picture was 
further elaborated: classically, at least, near $t=0$ the string
evolves as a set of weakly coupled ``string bits'', each of which
propagates smoothly  
across $t=0$. Assuming for now that this picture of the 
quantum transition makes sense, we can anticipate many of its main 
features. 
The production of excited string states may be calculated
using semiclassical methods obtaining a sensible
finite result \cite{turok}. The small parameter in 
the calculation turns out not to be 
$\alpha'$ but $\theta_0$, the rapidity 
of the orbifold plane collision: if $\theta_0$  is small a low
density of excited strings is produced. In \cite{niz} 
classical solutions describing strings passing through
$t=0$ have been studied in detail, showing 
how incoming strings undergo transmutation into a variety of 
excited massive states (see also \cite{tolley1}). 

One cannot expect the transition across $t=0$ to be describable 
by any effective theory including only massless fields. 
If effects involving the string coupling are neglected,
the excited states do not decay and they 
must appear in the final state.  
Neither the $\alpha'$ expansion nor any 
resummation of it involving the massless fields alone,
can possibly describe such a situation. 
Instead, a full string field theory approach involving the 
excited string states will likely be needed. 

Having made these necessary qualifications, we nevertheless want
to study the stringy $\alpha'$-corrections as 
the cosmic singularity is approached.  In the regime where it is valid, 
the $\alpha'$ expansion gives some
qualitative indications of the effects of virtual massive 
string states as 
$t=0$ approaches,  but before the production of real excited string 
states becomes significant. In particular, it is interesting to check
whether the correspondence between the M-theory background (\ref{metric11})
and the string theory background (\ref{metrics}) survives once
$\alpha'$-corrections are included, in the regime 
where the description of both theories in terms of the massless
 bosonic fields should still be valid. Here, the 
flatness of the M-theory background plays a critical role. 
As long as any higher order corrections 
can be expressed as powers of the Riemann curvature 
and the four-form field strength 
(recall there is no dilaton 
in eleven dimensions), none of them will have any 
affect on the linearized 
perturbations. Any correction they introduce in the perturbation
equations 
necessarily involves at least 
one power of the 
background curvature or field strength, both of which are zero. 
Therefore, the linearized perturbations continue to be described
by Einstein gravity 
on the M-theory side of the correspondence. 
All the $\alpha'$-corrections in string theory must therefore 
correspond to corrections in the map from M to string theory.
We shall explicitly construct such a covariant map,
order by order in $\alpha'$, and show that 
it consistently describes 
our results. 

While the technology for computing the $\alpha'$-corrections to the
bosonic background fields in string theory
has been much studied \cite{callan}-\cite{townsend}, several important ambiguities remain. 
The effective action for the massless fields can be computed either from
the tree-level string S-matrix \cite{callan} or from the nonlinear 
$\sigma $-model representing a string propagating 
in the relevant nontrivial background \cite{lovelace}. In
the second method, conformal (Weyl) symmetry of the quantum string
requires the vanishing of the $\beta$-functions in the 
$\sigma$-model. The latter are expressed as 
equations of motion for the massless bosonic fields, from which one
can reconstruct an effective action as a series of geometrical quantities at
each order in $\alpha'$. 

To lowest (zeroth) order in $\alpha'$, the effective action is just that for
general relativity plus the usual terms for the dilaton and the
antisymmetric tensor field. Different
string theories lead to different $\alpha'$-corrections, but
generically, to order $n$ in $\alpha'$, the corrections
involve products of the form
$\mathbf{R}^{m_1}(\mathbf{\nabla}\phi)^{2m_2}
(\mathbf{\nabla}^2\phi)^{m_3}\mathbf{H}^{2m_4}$ with 
$m_1+m_2+m_3+m_4=n+1$, 
where $\mathbf{R}$ is the Riemann tensor, $\phi$ is the dilaton and
$\mathbf{H}$ is the field strength of the antisymmetric tensor.
For type II theories, the 
$\mathcal{O}({\alpha'}^1)$ and $\mathcal{O}({\alpha'}^2)$ terms are zero, and
the first correction comes at order $\alpha'^3$ \cite{freeman}. 
In the case of the
bosonic or heterotic theories, however, there is a nontrivial
correction at first order in  
$\alpha'$ \cite{sen, sloan}. In this paper, we shall focus on the
heterotic string since it is the natural limit of heterotic M-theory
when the M-theory dimension becomes small, {\it i.e.} at weak string
coupling.

Generically, the $\alpha'$  corrections to Einstein's theory give field
equations with higher order time derivatives, possessing 
spurious solutions with bad physical behavior. 
As long as one 
is only interested in
perturbation
theory in $\alpha'$, such spurious solutions can be safely ignored: 
the higher order corrections are 
used only to correct the lower order solutions order
by order and no such problems are encountered. Nevertheless,
it is important to emphasize that to a given order in $\alpha'$,  
the action itself is ambiguous. 

In fact, for the first nontrivial correction,
one can remove the higher order time derivatives by adding certain
terms which vanish using the 
zeroth order equations of motion, as was shown
by Zwiebach \cite{zwiebach}. The point was generalized 
by Hull and Townsend \cite{townsend} who considered 
arbitrary covariant local field redefinitions such as 
\ba\label{redefinition}
g_{\mu\nu}'&=&g_{\mu\nu}+\alpha'(\mu_1R_{\mu\nu}+\mu_2
\nabla_\mu\phi\nabla_\nu\phi +\mu_3\nabla_\mu\nabla_\nu\phi+
g_{\mu\nu}(\mu_4R+ \mu_5(\nabla\phi)^2+\mu_6\nabla^2\phi)), \nonumber\\ 
\phi'&=&\phi+\alpha'(\mu_7R+\mu_8(\nabla\phi)^2+\mu_9\nabla^2\phi).
\ea
We have omitted terms involving the antisymmetric tensor field for
simplicity. Hull and Townsend 
showed that certain special linear 
combinations of these
$\mu$-terms do not alter the 
solutions to the equations of motion.
Such field
redefinitions lead to extra terms in the effective action which can be
written as squares of the lower order $\beta$-functions. 
If any other combination of the $\mu$'s is chosen,
then the resulting solutions to the equations of motion will be
different, but still 
physically equivalent because they correspond to
different definitions of the physical metric and dilaton field.
One can try to apply additional arguments
like the absence of higher time derivatives, duality
symmetry, and so on, in favor of certain choices,
but these are likely to remain inconclusive until the 
full result, including all orders in $\alpha'$, is known.

The $\alpha'$-corrected field equations for the massless 
string modes have been applied to a variety of 
different problems, from black hole thermodynamics 
(see e.g. \cite{blackholes}) to cosmology
\cite{applications}-\cite{akune}, and a number of 
interesting results have been obtained. Here, 
our intention is to use them to study 
the approach to a cosmological singularity, 
before the representation in terms of the massless fields alone
fails as we 
have argued it must.

The paper is divided as follows: in Section II we present the
effective action of the heterotic model. The next section describes
the cosmological background and its stringy corrections to second
order in $\alpha'$. Section IV shows how these stringy corrections
affect cosmological perturbations. We
present the tensor modes first and then the scalar perturbations where
some care must be taken with the gauge choice. Section V is devoted to
checking the proposed correspondence between M-theory and 
the heterotic string for this
particular background. In Section VI, we investigate the qualitative 
behavior of
the $\alpha'$-corrected solutions as they approach the singularity.
In the last section, we present some
conclusions. 

\section{Heterotic string effective action}

Our conventions follow the Landau-Lifshitz notation for the curvature
tensors, and we use a $(-,+,...,+)$ signature for the metric. Under
these assumptions, the form of the heterotic string effective action, in
string frame, including the first nontrivial $\alpha'$-correction but to
zeroth order in the string coupling, is \cite{lovelace,metsaev}
\be\label{action0}
S=\frac{1}{16\pi\, G}\int d^{10}x\sqrt{-g}\,e^{-2\phi}\Big(R+4(\nabla\phi)^2
+\frac{\alpha'}{8}R_{abcd}R^{abcd}+\mathcal{O}(\alpha'^2)\Big),
\ee
where $G$ is Newton's constant, and
$\phi$ is the dilaton. We shall not study the 
antisymmetric tensor contribution because it is zero for 
the background we are
interested in.

As discussed previously, the action (\ref{action0}) presents higher
derivative terms at the order $\alpha'$. However, after 
a particular field redefinition of the kind (\ref{redefinition}), with
$\mu_2=\mu_3=\mu_4=\mu_5=\mu_6=\mu_9=0$, $\mu_1=1$,
$\mu_7=1/8$ and $\mu_8=-1/2$, this effective
action takes the form \cite{sloan,zwiebach,townsend}
\ba
S&=&\frac{1}{16\pi\, G}\int
d^{10}x\sqrt{-g}\,e^{-2\phi}\bigg(R+4(\nabla\phi)^2 
+\frac{\alpha'}{8}\Big(R_{GB}^2+16G^{ab}\nabla_a\phi\nabla_b\phi-
16\nabla^2\phi(\nabla\phi)^2\nonumber \\ && \hspace{4.4cm}
+16(\nabla\phi)^4\Big)+ \mathcal{O}(\alpha'^2)\bigg), \label{action} 
\ea
which includes the Einstein's tensor $G_{ab}=R_{ab}-g_{ab}R/2$ and the
Gauss-Bonnet combination
$R_{GB}=(R_{abcd}R^{abcd}-4R_{ab}R^{ab}+R^2)$. Although not obvious, 
the present action yields second order field equations. 
However, the solutions
to the equations of motion remain the same as for the action
(\ref{action0}) if one constructs the solutions as a 
series in $\alpha'$ about the lowest order solution. 

The action (\ref{action}) has nice properties like the absence of physical
ghosts, unitarity and an $O(d,d)$ 
symmetry which is related to T-duality \cite{Osymmetry}. By analyzing string theory four point amplitudes in flat spacetime, Gross and
Sloan have worked out the quartic terms in the heterotic string action
up to third order in
$\alpha'$ \cite{sloan}. 
%Their $\alpha'$-correction includes all terms of relevance
%to our problem. 
They did not find any correction at $\alpha'^2$. They
did find a correction at $\alpha'^3$, but  
while this is sufficient for 
determining the $R^4$ term, for example, it does not 
include the coefficient of $(\nabla \phi)^8$ since 
that would require an eight point function. 
Unfortunately, the Gross-Sloan calculation is therefore incomplete 
for our purposes, since the dilaton is time-dependent in our background.
It would be interesting to complete their calculation, 
to further check the conjectured correspondence 
between the string and M-theory description of this 
background. In particular, 
as we shall argue below, there is an important relation which is
satisfied at 
lowest order by (\ref{metrics}), namely
$a^2 \propto e^{2\phi/ 3}$ where $a$ is the 
string frame scale factor. This relation
survives the first order correction in $\alpha'$. 
We have checked that it does {\it not} 
survive when Gross and Sloan's third order
correction is included, but that result remains 
inconclusive, as explained, until all the relevant 
contributions are included. 

In order to study solutions of the corrected action (\ref{action}) 
we proceed as follows.
The equations of motion are expressed as 
a series of ascending powers of $\alpha'$. The solution is then expressed 
as a series in $\alpha'$, in which the lower order terms enter 
the field equations as sources for the higher order terms. We first find the $\alpha'$-corrected background solution, and then
study the tensor and scalar
perturbations, using the $\alpha'$-corrected background.
Since the tensor modes are gauge
invariant, it is easier to start the discussion with them, and afterwards
we will focus on how to fix the gauge for the scalar modes to get
sensible results. 

\section{Cosmological background}\label{cosmobkg}

In this paper, we study the simplest case, where 
the string frame background is homogeneous, isotropic 
and flat in all nine space dimensions:
\be\label{metric10}
ds^2=a^2(t)\Big(-N^2(t)dt^2+\sum_{i=1}^9 (dx^i)^2\Big),
\ee
where $N(t)$ is the lapse function. 
We are also including a time-dependent dilaton $\phi$, but we shall 
set the antisymmetric tensor field to zero,
as mentioned
above. Now, we insert the background metric (\ref{metric10}) into the
action (\ref{action}) and obtain the equations of motion by taking
the variation with respect to the scale
factor, the dilaton and the lapse function
to get the Hamiltonian constraint.
The variation of (\ref{action}) with respect to the scale factor is 
\ba\label{vara}
0&=&288\,{a^6}\,{[(a^2)']}^2 - 576\,{a^8}\,(a^2)'\,\phi' +
160\,{a^{10}}\,{(\phi')}^2 + 288\,a^8\,(a^2)'' -  
144\,a^{10}\,\phi'' \nonumber \\&& + \alpha'\,\Big(
-189\,{[(a^2)']}^4 - 
504\,a^2\,{[(a^2)']}^3\,\phi' +  
1152\,a^4\,{[(a^2)']}^2\,{(\phi'(t))}^2 -
576\,a^6\,(a^2)'\,{(\phi')}^3  \nonumber \\ && 
\hspace{1cm}+80\,a^8\,{(\phi')}^4 +   
756\,a^2\,{[(a^2)']}^2\,(a^2)'' -
1008\,a^4\,(a^2)'\,\phi'\,(a^2)'' +
288\,a^6\,{(\phi')}^2\,(a^2)''  \nonumber \\&& 
\hspace{1cm}-504\,a^4\,{[(a^2)']}^2\,\phi'' +
576\,a^6\,(a^2)'\,\phi'\,\phi'' -
144\,a^8\,{(\phi')}^2\,\phi'' \Big),
\ea
where $'=\frac{d}{dt}$. The dilaton variation yields to
\ba\label{varb}
0 &=&
-72\,a^6\,{[(a^2)']}^2 + 128\,a^{8}\,(a^2)'\,\phi' -
32\,a^{10}\,(\phi')^2 - 72\,a^{8}\,(a^2)'' +  
32\,a^{10}\,\phi''  \nonumber \\&& 
+ \alpha'\,\Big( 63\,{[(a^2)']}^4 +
144\,a^2\,{[(a^2)']}^3\,\phi' -  
288\,a^4\,{[(a^2)']}^2\,(\phi')^2 +
128\,a^6\,(a^2)'\,(\phi')^3  
\nonumber \\&& \hspace {1cm}- 16\,a^8\,(\phi')^4 -  
252\,a^2\,{[(a^2)']}^2\,(a^2)'' +
288\,a^4\,(a^2)'\,\phi'\,(a^2)'' -
72\,a^6\,(\phi')^2\,(a^2)'' 
\nonumber \\&& \hspace {1cm}+  
144\,a^4\,{[(a^2)']}^2\,\phi'' -
144\,a^6\,(a^2)'\,\phi'\,\phi'' +
32\,a^8\,(\phi')^2\,\phi'' \Big),
\ea
and the lapse function constraint (after setting $N=1$) is
\ba\label{varc}
0 &=&
\left( 3\,(a^2)' - 2\,a^2\,\phi' \right) \,\bigg[ 48\,a^6\,(a^2)'
  - 16\,a^8\,\phi' +  
  \alpha'\,\Big( 63\,{[(a^2)']}^3 - 126\,a\,{[(a^2)']}^2\,\phi' 
  \nonumber \\&&\hspace{3cm}+
  60\,a^4\,(a^2)'\,{(\phi')}^2 -  
  8\,a^6\,{(\phi')}^3 \Big)  \bigg].
\ea 
To zeroth order in $\alpha'$, we obtain 
the Einstein-dilaton equations,
\ba
\nonumber 0 &=& 18\,(a^2)'+a^2\left[18\,(a^2)''+ a^2\left(
10\,{\phi'}^2 - 9\,\phi'' \right)  \right] - 36\,a^2(a^2)'\phi',
\\ \nonumber
0 &=& 9\,(a^2)'+ a^2\left[9\,(a^2)'' + 4\,a^2\left( {\phi'}^2 -
\phi''\right)  \right]  - 16\,a^2(a^2)'\phi',
\\ 
0 &=& \left(3\,(a^2)' -2 a^2\,\phi' \right)
\left(3\,(a^2)'-a^2\,\phi'\right).
\ea
which have the following solution:
\be\label{sol0}
a^2(t)=t, \qquad\qquad \phi(t)=\frac{3}{2}\ln(t).
\ee
We can interpret this solution by lifting it to eleven dimensions, using the
map
\ba\label{kkmetric}
ds^2_{11}&=&e^{4\phi/3}d\theta^2+e^{-2\phi/3}\,ds_{10}^2 \nonumber\\
&=& e^{4\phi/3}d\theta^2+e^{-2\phi/3}\,a^2[-dt^2+\sum_{i=1}^9(dx^i)^2],
\ea
which is simply compactified Milne times nine-dimensional flat space,
(\ref{metric11}). There is a second inequivalent solution with $a^2=
e^{\phi/3}$, in which the string frame scale factor and the
string coupling diverge as the singularity is approached. This
solution does not correspond to a flat background in eleven dimensions
so we shall not study it here.

In order to find the stringy corrections to the cosmological 
background we express them as a series in $\alpha'$, namely
\be
a^2(t)=a_0^2(t)+\alpha'a_I^2(t)+\alpha'^2a_{II}^2(t)+\cdots\quad\quad
\phi(t)=\phi_0(t)+\alpha'\phi_I(t)+\alpha'^2\phi_{II}(t)+\cdots
\ee
where $a^2_0(t)$ and $\phi_0(t)$ are given by the zeroth order solutions
(\ref{sol0}). To first order, the equations for $a_I$ and $\phi_I$ are then
sourced by the $\alpha'$-correction terms evaluated using the
zeroth order solutions.
Then, to first order in $\alpha'$ equations (\ref{vara})-(\ref{varc})
reduce to 
\ba
0&=&9 + 48\,t^2\,a_I^2 - 48\,t^3\,{(a_I^2)}' - 16\,t^4\,\phi_I' +
48\,t^4\,{(a_I^2)}'' -24\,t^5\,\phi_I'', \nonumber
\\0&=& \nonumber
9 + 24\,t^2 (a_I^2) -24\,t^3\,{(a_I^2)}' -
16\,t^4\,{\phi_I}' + 36\,t^4\,{(a_I^2)}''-16\,t^5\,\phi_I'',
\\0&=&
\frac{3\,{a_I^2}}{t} - 3\,{(a_I^2)}' + 2\,t\,\phi_I',
\ea
which are solved by
\be\label{firstcorr}
a_I^2(t)=-\frac{1}{8\, t^2}+c_1t+c_3, \qquad\qquad
\phi_I(t)=-\frac{3}{16\,t^3}+c_2-\frac{3c_3}{2t}.
\ee
The integration constants $c_1$, $c_2$ and $c_3$ can be removed by rescaling
and shifting time, and by a shift of the dilaton, respectively. Thus, the
only nontrivial corrections come from the first term of each
field. The negative sign of these terms is interesting: 
it means that the $\alpha'$-corrections act to strengthen 
the onset of the singularity. As we briefly discuss in the conclusions,
this is actually consistent with expectations based on the string
bits picture. One does not expect the universe to ``bounce'' in 
the field theory description. Rather, one expects 
a phase transition which cannot be described in general relativity.

As mentioned, the calculations of 
\cite{sloan} found no corrections to the effective action 
at order $\alpha'^2$. Assuming all such terms are ruled out,
the second order correction to the background solution may be
computed 
using only the
first order $\alpha'$-correction in the action. 
At second order in $\alpha'$, 
equations (\ref{vara})-(\ref{varc}) read:
\ba
0&=& 153 + 768\,t^7\,(a_{II}^2)' - 256\,t^7\,\phi_{II}' +
768\,t^8\,(a_{II}^2)'' - 
384\,t^8\,\phi_{II}'', \nonumber
\\0&=& \nonumber
27 + 96\,t^7\,(a_{II}^2)' - 32\,t^7\,\phi_{II}' + 72\,t^8\,(a_{II}^2)''-
32\,t^8\,\phi_{II}'',
\\0&=&
9 + 192\,t^7\,(a_{II}^2)' -128\,t^7\,\phi_{II}',
\ea
and their solution is 
\be\label{secondcorr}
a_{II}^2(t)=-\frac{9}{160\, t^5}+c_1t+c_3, \qquad\qquad
\phi_{II}(t)=-\frac{123}{1280\,t^6}+c_2-\frac{3c_3}{2t}.
\ee
Again the integration constants $c_1$, $c_2$ and $c_3$ can be removed
by coordinate transformations. One can similarly calculate the
$\alpha'^3$ corrections coming from the Gross-Sloan action, even though
this is subject to the caveat made in the introduction. In fact, the
coefficient of $\alpha'^3$ in Gross and Sloan's correction term is
small, and the leading correction to the solution at order $\alpha'^3$
actually comes from the the first order $\alpha'$-correction in the
action. The latter correction is  
\be\label{thirdcorr}
a_{III}^2(t)=-\frac{63}{1280\, t^8}, \qquad\qquad
\phi_{III}(t)=-\frac{437}{5120\,t^9},
\ee
where we have removed the integration constants. 

\begin{figure}[t!]
{\centering
\resizebox*{3in}{3in}{\includegraphics{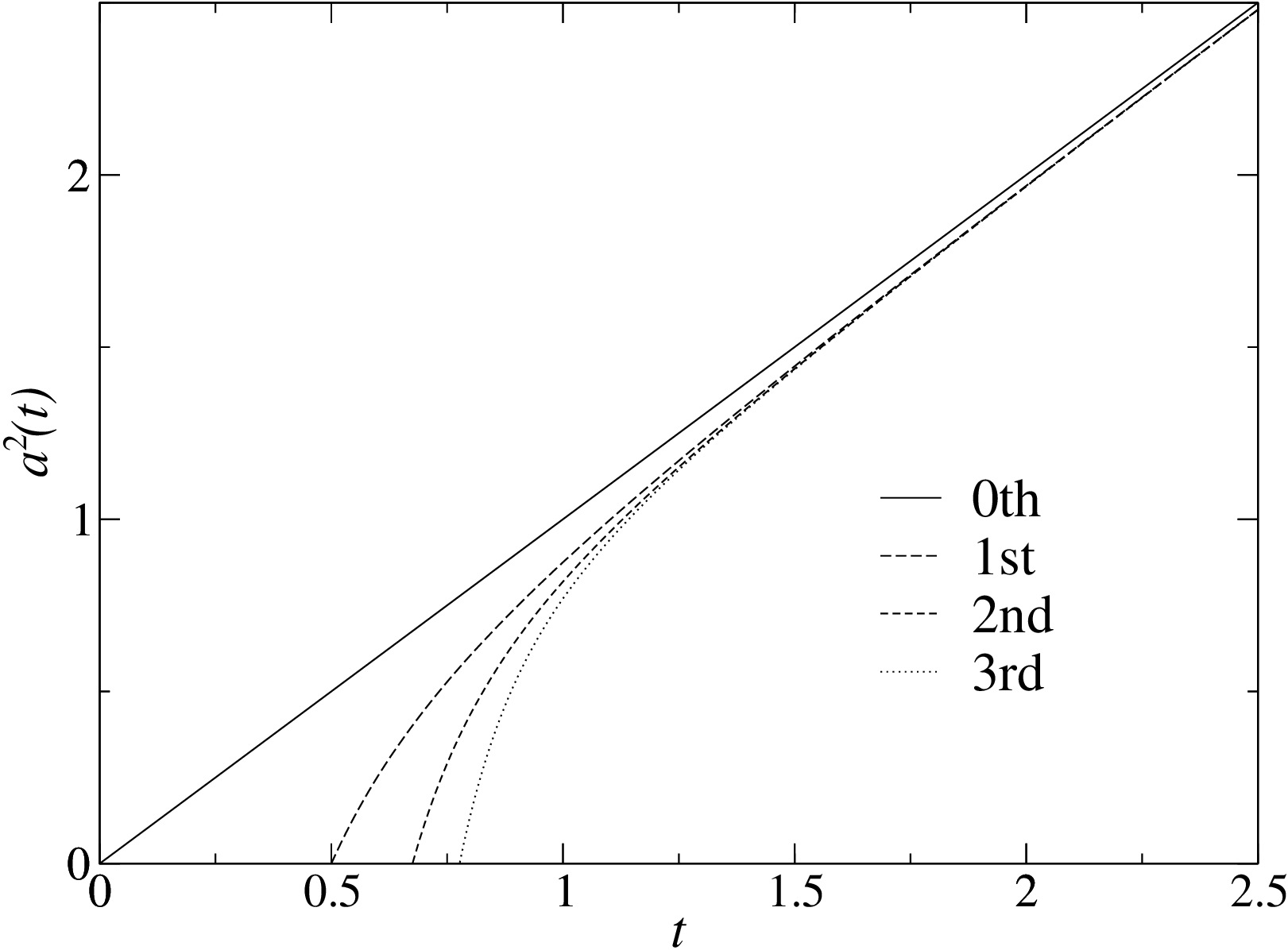}}
\resizebox*{3in}{3in}{\includegraphics{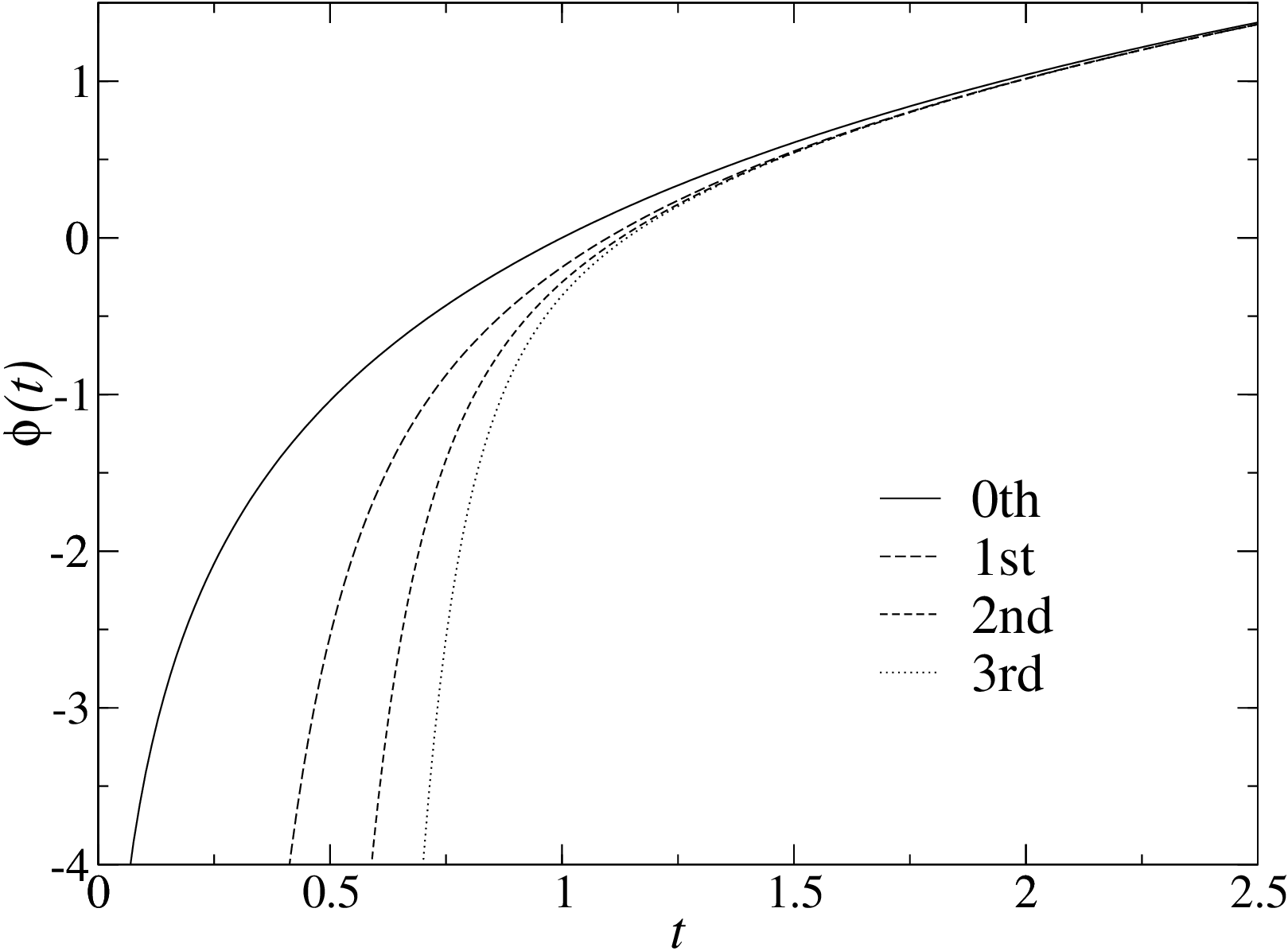}} }
\caption{The scale factor $a^2(t)$ and the dilaton
  $\phi(t)$ with $\alpha'$-corrections in units of
  $\alpha'=1$. The solid line is the zeroth order solution and as we
  move from left to right we get the first, second and third order 
  $\alpha'$-corrections. 
  } 
\label{bkground}
\end{figure}

Notice that the corrections only become significant 
for $t<(\alpha')^{1/3}$, the string time, consistent with
naive expectations (see Figure \ref{bkground}). 
Furthermore, the corrections act to  
strengthen the singularity in the scale factor and the 
divergence of the dilaton to $-\infty$. 
As shall be explained 
in the conclusions, this is actually consistent with the
string bits picture.

%% Another interesting result is that if we expand $3/2 \ln(a^2)$ as a series
%% in $1/t$ then we recover the dilaton's solution, including the
%% $\alpha'$-corrections, namely
%% \be
%% \frac{3}{2} \ln (a_0^2+\alpha'a_{I}^2+\alpha'^2a_{II}^2)=
%% \frac{3}{2}\ln(t)-\frac{3\,\alpha'}{16\,t^3}-
%% \frac{123\,{\alpha'}^2}{1280\,t^6}+{\cal
%%   O}\left(\frac{1}{t^9}\right)\simeq\phi(t). 
%% \ee
%% Equivalently, we could have chosen to expand $e^{2/3\phi}$
%% to obtain an expression equal to the scale factor. To our
%% understanding this shows the geometrical origin of the
%% dilaton in eleven dimensions, though there is not evidence that this
%% should hold to all orders in $\alpha'$.

\section{Cosmological Perturbations}

We now turn to calculating the behaviour of linearized  perturbations about 
this cosmological background (\ref{metric10}). 
The perturbations can, as usual, be decomposed 
into a transpose-traceless tensor component, a vector piece and
a scalar degree of freedom, which each evolve independently. 
Here, we shall ignore the vector perturbations since there are no vector sources at the linearized level. 

\subsection{Tensor Perturbations}

We begin with the tensor perturbations, which are gauge invariant and hence 
technically the easiest to study. The general tensor perturbation may be expressed as a linear combination of gravitational plane waves 
and, without loss of generality, we can focus on just one wave 
propagating in the $\hat{k}$ direction, {\it i.e.}
$h_{ij} \propto
e^{i \vec{k}\cdot \vec{x}}$, where 
\be\label{tensormetric}
ds^2=a^2(t)\Big(-dt^2+\sum_{j=1}^9\sum_{i=1}^9
(\delta_{ij}+h_{ij})dx^idx^j\Big),
\ee
and for tensor perturbations,
$\sum_j h_{ij}k_j=\sum_j h_{ji}k_j=h_{ii}=0$, with
$|h_{ij}|\ll 1$.
To zeroth order in $\alpha'$, the
variation of action (\ref{action}) with respect to the tensor perturbation 
gives
\be
{h}''_{ij}+\frac{1}{t}{h}'_{ij}+k^2h_{ij}=0,
\ee
where $'=d/dt$ as before.
The general solution is given in terms of the Bessel functions: 
\be
h_{ij}=A_{ij}J_0(kt)+\frac{\pi}{2}B_{ij}K_0(kt),
\ee
where $A_{ij}$ and $B_{ij}$ are constant
matrices, and the factor $\frac{\pi}{2}$ is convenient when expanding for
small $k\equiv |\vec{k}|$. The $J_0$ function is regular at $t=0$ whereas 
$K_0$ is logarithmically divergent. We shall mainly be interested 
in the behavior of the perturbations at long wavelengths,
$k\rightarrow 0$, for which the solution reduces to 
\be\label{pert0}
h_{ij}={A}_{ij}+ B_{ij}\ln\left(\frac{kt\,
  e^\gamma}{2}\right), 
\ee
where Euler's constant $\gamma\simeq 0.5772$. Note that
$h_{ij}$ has a trivial regular piece proportional to $A_{ij}$, and an
irregular component given by the $B_{ij}$ term, which diverges as
$t\rightarrow 0$. Furthermore, note that the solution tends to
minus infinity logarithmically as $k\rightarrow 0$. 

\begin{figure}[t!]
{\centering
\resizebox*{3in}{3in}{\includegraphics{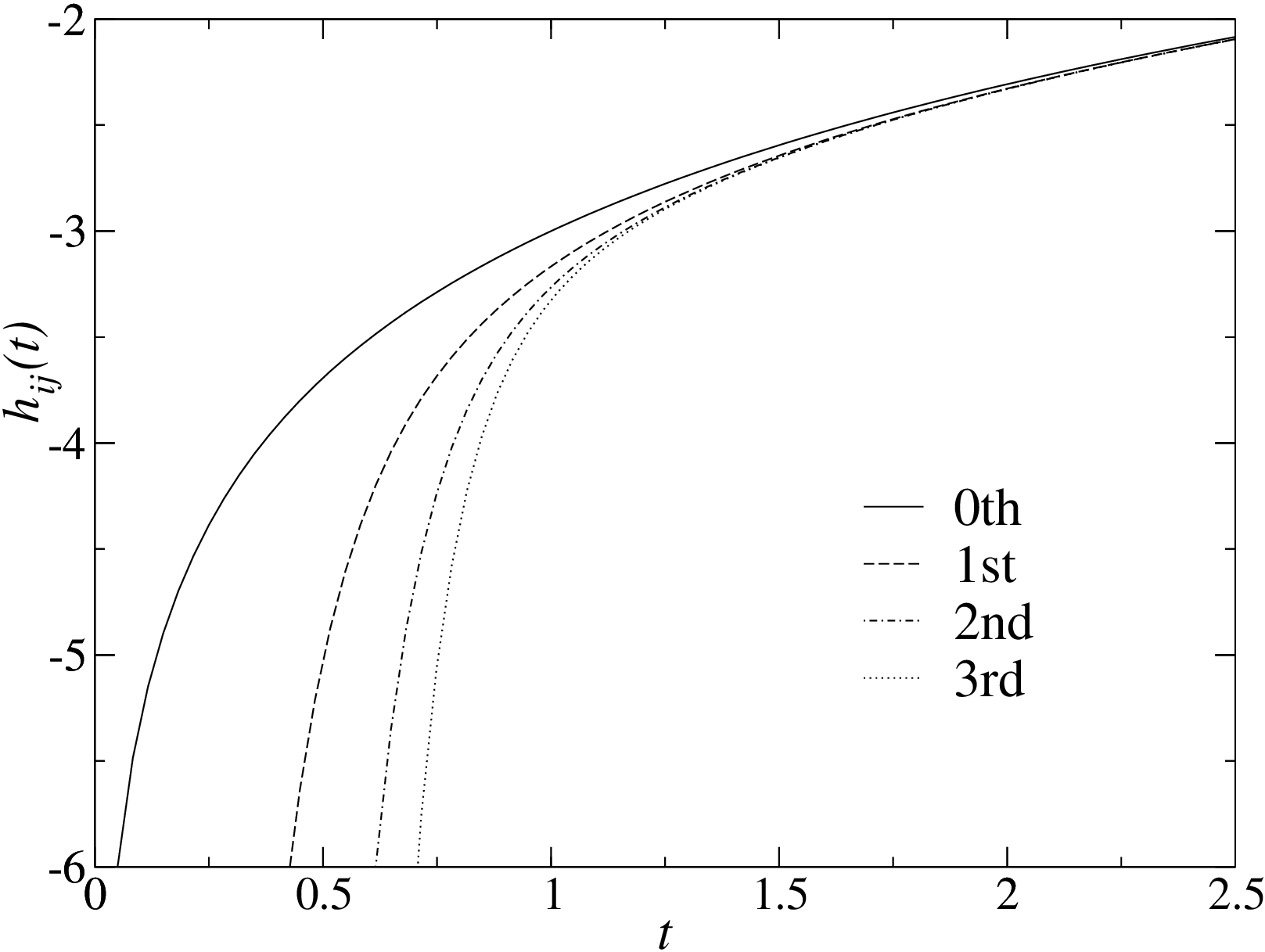}} 
%\resizebox*{3in}{3in}{\includegraphics{pert2.eps}} 
}
\caption{The plot shows a single component of the tensor perturbation
  $h_{ij}$ in the limit of small $k$ as a function of time, focusing on its
  behavior near $t=0$. The zeroth order solution is the solid line and
  as we move to the right we get the first, second and third order
  corrections respectively. We have chosen $A_{ij}=-3$ and $B_{ij}=1$,
  which corresponds to $k\sim 0.02$ in the solution with
  non-zero $\vec{k}$, given by equation (\ref{pert0}), with
  $A_{ij}=B_{ij}=1$.} 
\label{tensorpert}
\end{figure}

Through a similar analysis to that explained above for the background, 
we can 
compute the $\alpha'$-corrections to the tensor perturbations. 
First we express the $h_{ij}$ as a series in ascending powers of $\alpha'$,
$h_{ij} = h_{ij}^0 +h_{ij}^I\alpha' +h_{ij}^{II} \alpha'^2 +\dots$, 
starting with the above zeroth order solution. Then we solve the 
equations of motion including $\alpha'$-corrections to the background, 
order by order in $\alpha'$, for the tensor perturbation 
at each successive power
of $\alpha'$. The equation for the 
first order correction $h_{ij}^I$ reduces to 
\be\label{tensoreqn}
({h}^{I}_{ij})''+\frac{1}{t}({h}^{I}_{ij})'+k^2h^I_{ij}
 =   \frac{1}{4\, t^3}\,\left( k^2\,h^0_{ij} +
4\,(h^0_{ij})''\right)- \frac{1}{2\,t^4}(h^0_{ij})',
\ee
where the LHS is just the Bessel differential operator acting on
$h^I_{ij}$, and the RHS is the source term given at small $k$ 
in terms of the zeroth order solution $h^0_{ij}$ (\ref{pert0}). Solving at
small $k$, we obtain 
\be
h_{ij}^I=-\frac{B_{ij}}{6\,t^3}+(d_1)_{ij}+(d_2)_{ij}\ln(t),
\ee
where the integration constants $(d_1)_{ij}$ and $(d_2)_{ij}$ can
be absorbed into $A_{ij}$ and $B_{ij}$. Thus the only non-trivial
correction comes from the term proportional to $t^{-3}$.
This term has the same negative sign we found for the 
background, and as before, its effect is to strengthen the singular behavior
of the solution. The coefficient is again small, so that 
the corrections only become important at around $t\sim
1$ in units of $\alpha'=1$ (see Figure \ref{tensorpert}). 

To second order in $\alpha'$ we obtain a similar equation
for $h^{II}_{ij}(t)$, sourced by the zeroth and first order solutions, namely
\be\label{tensoreqn2}
({h}^{II}_{ij})''+\frac{1}{t}({h}^{II}_{ij})'+k^2h^{II}_{ij}
 = \frac{k^2\,{h_{ij}^0}}{32\,t^6} +
 \frac{k^2\,{h^I_{ij}}}{4\,t^3} -
 \frac{169\,{(h_{ij}^0)}'}{160\,t^7} -    
  \frac{{(h^I_{ij})}'}{2\,t^4} + \frac{7\,{(h_{ij}^0)}''}{32\,t^6} +
  \frac{{(h^I_{ij})}''}{t^3}.
\ee
It is again simple to solve this in the limit of small $k$, 
obtaining 
\be
h_{ij}^{II}=-\frac{47\,B_{ij}}{480\,t^6}+(d_1)_{ij}+(d_2)_{ij}\ln(t),
\ee
where the integrations constants can again be absorbed in the zeroth
order solution.

One can straightforwardly repeat the process for nonzero $k$, obtaining the solution as a series in $k t$: 
\ba\label{nonzerok}
h_{ij}&=& h_{ij}^0 \nonumber
\Bigg[1-\frac{(kt)^2}{4}-\frac{(kt)^4}{64}+...
  -\frac{\alpha'}{t^3}\bigg(\frac{13(kt)^4}{32}+...\bigg)
  +\frac{\alpha'^2}{t^6}\bigg(\frac{9(kt)^2}{320}+
  \frac{251(kt)^4}{2560}+...\bigg)
  \Bigg] \\ &&\nonumber
+B_{ij} \Bigg[ \frac{(kt)^2}{4}-\frac{3(kt)^4}{128}+...
  -\frac{\alpha'}{6t^3}\bigg(1 
  + \frac{29(kt)^2}{4}+\frac{653(kt)^4}{64}+...\bigg)\\ &&
  \hspace{1.2cm}+ \frac{\alpha'^2}{96t^6}\bigg(\frac{47}{5}+
  \frac{547(kt)^2}{20} +
  \frac{4481(kt)^4}{64}+...\bigg)
  \Bigg],
\ea
where $h^0_{ij}$ is the zeroth order solution 
(\ref{pert0}) and the dots represents terms of order 
${\cal O}(k^6)$. At finite $k$ we see that even the regular $A_{ij}$ mode
eventually suffers divergent corrections near $t=0$. This is just 
as one expects. 
When the physical wavelength of a mode, $L=t^{1\over 2} /k$, becomes
smaller than the string scale, $L_s\equiv \sqrt{\alpha'}$, the 
$\alpha'$-corrections become large. All of the corrections 
may be expressed as positive powers of $L/T$ and $L_s/T$ 
where $T$ is the proper time in string frame, $T ={2\over 3} t^{3\over 2}$. 

We have also computed the correction at order $\alpha'^3$, coming from
the $\alpha'$ term in the effective action and also from
Gross and Sloan's correction at $\alpha'^3$. For the former we find
\be
h_{ij}^{III}=-\frac{2183\,B_{ij}}{23040\,t^9}.
\ee
We have checked that the Gross-Sloan correction is much 
smaller than this. But since that term is in any case incomplete,
we shall not bother to state the correction here.

The behavior of the corrected solution up to $\alpha'^3$ 
is shown in Figure
\ref{tensorpert}. Now, let us turn to a consideration of 
the scalar
perturbations.

\subsection{Scalar Perturbations}

The most general scalar perturbation about our background solution
can be 
written as 
\be\label{metricscalar}
ds_{10}^2 =a^2(t)[-(1+2\Upsilon_{10})dt^2-
  2\partial_i\Omega_{10}\, dx^idt+[(1-2\Psi_{10})\delta_{ij}-
    2\partial_i\partial_j\chi_{10}]dx^idx^j]. 
\ee
It will be useful to keep in mind the Kaluza-Klein map to eleven 
dimensions, which encodes the geometrical role of the
dilaton:
\be\label{11dmap}
ds^2_{11}=e^{\frac{4}{3}(\phi+\delta\phi)}d\theta^2+
e^{-\frac{2}{3}(\phi+\delta\phi)}\,ds_{10}^2.
\ee
The equations of motion for the perturbations are calculated 
by expanding the action (\ref{action}) to second order in
the perturbations and calculating the resulting 
Euler-Lagrange equations. In order to calculate the perturbations
we have to fix a gauge, but it is useful to identify gauge invariant
quantities in terms of which we can express the results.
Since all perturbations have the same origin, as gravitational
waves in flat eleven-dimensional spacetime, 
we expect to be able to choose coordinates
so that they all obey the same equations of motion. We
shall show that by choosing the gauge appropriately in
the scalar sector, all the scalar perturbation variables 
end up obeying exactly the same equations as the tensor
perturbations, up to the order in $\alpha'$ which we work. 

Under the change of coordinates $x^\mu\rightarrow x^\mu+\xi^\mu$,
with $\xi^\mu$ small, it is straightforward to check that the 
scalar perturbations transform as 
\ba
\Upsilon_{10}&\rightarrow& \Upsilon_{10}-\partial_0\xi^0-
\frac{a'}{a}\xi^0,\\ 
\delta\phi&\rightarrow& \delta\phi-\phi'\xi^0,\\
\chi_{10}&\rightarrow& \chi_{10}+\xi^s,\\
\Psi_{10}&\rightarrow& \Psi_{10}+
\frac{a'}{a}\xi^0,\\
\Omega_{10}&\rightarrow& \Omega_{10}-\xi^0+\partial_0\xi^s,
\ea
where $\xi^i \equiv \partial_i \xi^s$. 
In particular, we notice that
\be\label{rho}
\rho=\Psi_{10}+\frac{1}{3}\delta\phi
\ee
is a gauge-invariant quantity, because $a'/a= \phi'/3$. 
In fact, $\rho$ is just the eleven-dimensional isotropic
perturbation 
$\Psi_{11}$, which is gauge invariant just because the spatial metric
is static in eleven dimensions. As we shall discuss later, we expect
the condition $a^2 = e^{2\phi/3}$, necessary for this correspondence
to hold, to be enforcible by suitable field redefinitions 
to all orders in $\alpha'$.

Now, to fix the gauge, it is convenient to make the choice that
the  spatial metric perturbation to the eleven-dimensional metric
be traceless. This condition ensures that in the long wavelength
limit, the solutions are linearized versions of the well-known
Kasner solutions. In terms of our ten dimensional variables,
this condition reads:
\be
\chi_{10}=\frac{9}{k^2}\left(\Psi_{10}+\frac{1}{3}\delta\phi\right),
\ee
and it fixes the $\xi^s$ gauge freedom completely. One remaining gauge
choice is needed to fix $\xi^0$. A second relation is then 
found from the field
equations. By adjusting the gauge condition order by order in 
$\alpha'$, we find that the conditions
\ba\label{gauge}
\Upsilon_{10}&=&-4\left(1-\frac{3\,\alpha'}{(2\,t)^3}-\frac{3(5\,
  \alpha')^2}{(2\,t)^6}  \right)\Psi_{10},\nonumber\\
\Psi_{10}&=&-3\left(1+\frac{\alpha'}{2\,
  t^3}+\frac{13\,{\alpha'}^2}{16\, t^6}\right)\rho,
\ea
result in an equation for $\rho$ which is
exactly the same equation as that found earlier for 
the tensor perturbation $h_{ij}$, for all $k$.

It is helpful to re-interpret the result in eleven dimensions. The
general scalar perturbation of (\ref{metric11}) involving the lowest Kaluza-Klein modes, and no gauge fields (these are projected out by the $Z_2$ orbifolding in the heterotic theory) is 
\be\label{scalar11}
ds^2_{11}=-(1+2\Upsilon_{11})dt^2+t^2(1-2\Gamma_{11})d\theta^2-
2\partial_i\Omega_{11} \,dx^idt+[(1-2\Psi_{11})\delta_{ij}-
  2\partial_i\partial_j\chi_{11}]dx^idx^j.
\ee
Comparing with (\ref{metricscalar}) and (\ref{11dmap}), we see that 
\ba
\Gamma_{11}&=& -{2\over 3} \delta \phi,\label{gamma11}\\
\Psi_{11}&=&\Psi_{10}+\frac{1}{3}\delta\phi=\rho,\\
\Upsilon_{11}&=&\Upsilon_{10}-\frac{1}{3}\delta\phi.
\ea
To lowest order in $\alpha'$, the $(t,\theta)$ part of
the eleven-dimensional metric
is only conformally perturbed in this gauge
 ({\it i.e.} $\Gamma_{11}= -\Upsilon_{11}$). This is a nice feature
in providing a geometrical interpretation of the matching conditions
across the bounce, as explained in \cite{McFadd}. 
This property is spoiled
by the higher order $\alpha'$-corrections, if one adopts the naive 
map (\ref{11dmap}). However, as we shall see later, once the map is
suitably adjusted, this feature of the eleven dimensional metric is retained.

As mentioned, the field $\rho$ obeys the same equation as 
the tensor modes, for all $k$, confirming the idea of a common
higher dimensional origin. In the limit of small $k$, the solution 
is thus
\be\label{psi11sol}
\rho=  A + B\,\ln (t)-\frac{\alpha'\,B}{6\,t^3} -
\frac{47\,{\alpha'}^2\,B}{480\,t^6} +\dots
\ee
For nonzero $k$ we get (\ref{nonzerok}), 
but with $A$ and $B$ instead of $A_{ij}$
and $B_{ij}$. We can calculate the dilaton perturbation
using (\ref{gauge})
and the solution for $\rho$. In the limit of small $k$, this is
\be
\delta\phi=(A+B\ln(t))\left(12+\frac{9\,\alpha'}{2\,t^3}+
\frac{117\,\alpha'^2}{16\,t^6}\right)-\frac{B\,\alpha'}{t^3}\left(2
+\frac{77\,\alpha'}{40\,t^3}\right). 
\ee
If the eleven dimensional picture is correct, we expect it is possible to
choose a gauge in which scalars and tensors obey the same 
equation, to all orders in $\alpha'$. 

\section{Eleven dimensional connection}

At lowest order in $\alpha'$, we have the relation $a^2= e^{2 \phi/3}$, and
as we have seen, this is preserved at first and second order in
$\alpha'$. There are reasons to believe the relation will continue to
hold, perhaps after a suitable field redefinition, to all orders. The
reason is that in the eleven dimensional picture of M-theory, the
gauge fields live on the two orbifold planes which, in the solution
(\ref{metric11}) neither expand nor contract. This is consistent with
the gauge field Lagrangian in heterotic string theory, which takes the form
$\int d^{10} x \sqrt{-g} e^{\lambda \phi} F^2$, with $\lambda$ 
such that the dilaton's time dependence precisely cancels that of the
scale factor, if the relation $a^2= e^{2 \phi/3}$ is
satisfied. Therefore, we can expect this relation to hold to all
orders, possibly after a field redefinition in string frame, 
if the metric (\ref{metric11}) is a solution of M-theory.

In this section, we want to go further and construct a map from eleven to
ten dimensions. Since the eleven dimensional metric should receive no corrections, all of the $\alpha'$-corrections have to arise from the map. We
will consider, in order, the background, perturbations at long wavelengths,
and then the leading nontrivial $k$-dependence.
The most general covariant expression for the ten dimensional string frame metric in terms of the eleven dimensional M-theory metric is
\ba\label{map}
g_{\mu\nu}^{(10)}dx_{10}^\mu dx_{10}^\nu&=&e^\gamma\left(1+m_0\alpha'
e^{-\gamma}(\nabla\gamma)^2+...\right)dx_{11}^\mu dx_{11}^\nu
\bigg(g_{\mu\nu}^{(11)} 
+m_1\alpha'e^{-\gamma}\nabla_\mu\gamma\nabla_\nu\gamma
\nonumber \\ && 
+m_2\alpha'e^{-2\gamma}\nabla_\mu(\nabla_\nu e^\gamma)
+m_3\alpha'e^{-\gamma}R_{\mu\nu}^{(11)}
+m_4\alpha'e^{-\gamma}g_{\mu\nu}^{(11)}R^{(11)}+...\bigg)
,
\ea
where $\mu,\nu$ run over the ten string-theory dimensions and 
$\gamma={1\over 2} \ln(g^{(11)}_{\theta\theta})$, and the ellipses indicate
higher order $\alpha'$-corrections. Since we are working at long
wavelengths, we restrict ourselves to terms with only 
two derivatives.
The form of each term and, in particular, the 
powers of $e^\gamma$ are determined by dimensional
analysis. 

We  compute the RHS of 
(\ref{map}) using the eleven dimensional metric 
(\ref{metric11}) with tensor or scalar perturbations. We then compare it
with the LHS using the ten dimensional results from the $\alpha'$ 
expansion, and attempt to fix the coefficients
$m_a$ ($a=0,...,4$).
If we insist on the relation $a^2=e^{2\phi/3}$, we have
\be\label{dilconf}
e^{2\phi/3}=e^\gamma\left(1+
m_0\alpha'e^{-\gamma}(\nabla\gamma)^2+...\right). 
\ee

In the background (\ref{metric11}), then
$\gamma=\ln(t_{11})$ and the spatial components of the map
(\ref{map}) give 
\be
a_{(10)}^2\delta_{ij}
\ee 
for the LHS, and 
\be
e^\gamma\left(1+
m_0\alpha'e^{-\gamma}(\nabla\gamma)^2+...\right)\delta_{ij}
\ee
for the RHS, consistent by construction with $a^2=e^{2\phi/3}$. 
However, the time-time component of (\ref{map}) leads to 
\be
-a^2=(dt_{11}/dt)^2 t_{11}
(1+m_0\alpha'/t_{11}^3)(-1+m_1/t_{11}^3),
\ee
where $t=t_{10}$. After expanding to first order in $\alpha'$ and
integrating both sides of the map, we find that the eleven dimensional
time is related to ten dimensional time by
\be\label{timetrans}
t_{11}=t\left(1-m_1\frac{\alpha'}{4\,
  t^3}+\mathcal{O}(\alpha'^2)\right),
\ee
since the term involving $m_0$ cancels. 
Therefore, the time in eleven dimensions is not the same as in ten
and, in particular, the singularity at $t_{11}=0$ gets 
mapped to a positive time in string frame for $m_1>0$. 

To complete the background analysis, we need to 
compare the dilaton expansion (\ref{dilconf}) with its stringy
corrections (\ref{firstcorr}). After rewriting the stringy corrections
using the eleven dimensional time $t_{11}$, we get a relation for the
unknown coefficients $m_0$ and $m_1$, given by
\be\label{condition1}
m_0+\frac{1}{4}m_1=\frac{1}{8}.
\ee

Turning to the tensor perturbations, the only
piece of the map that changes is the tensor component in the metric
$g_{ij}^{(11)}$: the LHS of the map 
(\ref{map}) reads
\be
a^2(\delta_{ij}+h_{ij})=a^2\left[\delta_{ij}+A_{ij}+B_{ij}\left(\ln(t)-
  \frac{\alpha'}{6 t^3}\right)\right],
\ee
and the RHS reads
\be
e^{2\phi/3}\left(\delta_{ij}+h_{ij}^{(11)}+\frac{m_2\alpha'}{2\,t_{11}^2}
\Gamma^0_{ij}\right)=  
e^{2\phi/3}\left[\delta_{ij}+A_{ij}+B_{ij}\left(\ln(t_{11})-\frac{m_2\alpha'}{ 
    t_{11}^3}\right)\right].
\ee
By comparing both sides and using the time transformation
(\ref{timetrans}), we get another constraint for the unknown
coefficients $m_1$ and $m_2$, 
\be\label{condition2}
m_2+\frac{m_1}{2}=\frac{1}{3}.
\ee

In the case of the scalar perturbations the comparison between the
LHS and RHS of the map (\ref{map}) is not as simple as for the tensor
modes, due to the fact that we have to take into account the perturbation
in $\gamma$, which is given by $\gamma\rightarrow
\gamma+\delta\gamma=\ln(t)(1-\Gamma_{11})$, and, more importantly,
there is a remaining gauge freedom.
Thus one has to be careful to choose a gauge invariant
variable to compare on both sides of the map. As we have seen in the
last section, the quantity $\rho$ defined in (\ref{rho}) is one
such quantity, which is furthermore 
local in time. This means we 
do not need to use any of the time
components of the map (\ref{map}), which would have 
involved an
integration, just as occurred for the background.
Under the gauge choice (\ref{gauge}), the
quantity $\rho$ simplifies to 
\be\label{rho1}
\rho= A+B\left(\ln(t)-\frac{\alpha'}{6\,t^3}\right)+
\mathcal{O}(\alpha'^2).
\ee
Using the RHS of the map (\ref{map}) and 
the dilaton expansion (\ref{dilconf}), we get the following
expressions for $\Psi_{10}$ and $\delta\phi$ in terms of 11d
quantities 
\ba
\Psi_{10}&=&-3\left(1+\alpha'\frac{4\,m_0}{t_{11}^3}\right)\Psi_{11}
+\alpha'\frac{16\,m_0-m_2}{2\,t_{11}^2}\partial_0\Psi_{11},\nonumber \\
\delta\phi&=&12\left(1+\alpha'\frac{3\,m_0}{t_{11}^3}\right)\Psi_{11}
-\alpha'\frac{24\,m_0}{t_{11}^2}\partial_0\Psi_{11},
\ea
which results in
\be\label{rho2}
\rho\equiv\Psi_{10}+\frac{\delta\phi}{3}=\Psi_{11}
-\alpha'\frac{m_2}{2\,t_{11}^2}\partial_0\Psi_{11}=A+B\left(\ln(t_{11})-
\alpha'\frac{m_2}{2\,t_{11}^3}\right).
\ee
By comparing both equations (\ref{rho1}) and (\ref{rho2}), we get
again condition (\ref{condition2}). Therefore, the map is
consistent. 
The terms involving the
Ricci tensor and Ricci scalar do not contribute at all to this background,
leaving $m_3$ and $m_4$ unfixed; and the two conditions (\ref{condition1})
and (\ref{condition2}) can only fix two of the three remaining
coefficients. In particular, we can take $m_0=0$ and
then fix the other two parameters to be: $m_1=1/2$ and $m_2=1/12$. This
choice represents a simple and physical picture: if $m_0=0$
then $e^{2\phi/3}= t_{11}$, which implies that the dilaton
really measures the eleven-dimensional distance between the two orbifold planes.

Finally, let us see how the map can 
be extended to describe perturbations at finite $k$. We focus on
tensor perturbations because of their gauge invariance. 
We include terms which only contribute to a desired order in $k$ and
which do not change the mapping for the background or at lower
orders
in $k$. At order $k^2$,
we have found two terms which provide two 
unknown parameters needed to recover the 10d result
(\ref{nonzerok}). The map with these new extra terms is 
\ba\label{map2}
g_{\mu\nu}^{(10)}dx_{10}^\mu dx_{10}^\nu&=&e^\gamma\left(1+m_0\alpha'
e^{-\gamma}(\nabla\gamma)^2+...\right)dx_{11}^\mu dx_{11}^\nu
\bigg(g_{\mu\nu}^{(11)} 
+m_1\alpha'e^{-\gamma}\nabla_\mu\gamma\nabla_\nu\gamma
 \\ && \nonumber\!\!\!
+m_2\alpha'e^{-2\gamma}\nabla_\mu(\nabla_\nu e^\gamma)
+m_3\alpha'e^{-\gamma}R_{\mu\nu}^{(11)}
+m_4\alpha'e^{-\gamma}g_{\mu\nu}^{(11)}R^{(11)}\\ && \!\!\!\nonumber
+m_5\alpha'(\nabla^\lambda e^{-\gamma})\nabla_\lambda\Big(e^{4\gamma}R_{\mu\  
  \nu}^{\ \alpha\ \beta}\nabla_\alpha\gamma\nabla_\beta\gamma\Big)\!
+m_6\alpha'(\nabla^\lambda e^{-\gamma})\nabla_\lambda\Big(e^\gamma \nabla_\mu\nabla_\nu 
e^\gamma\Big) \!
+...\bigg),
\ea
where $R_{\mu\ \nu}^{\ \alpha\ \beta}$ is the 11d Riemann tensor.
We calculate the RHS of this map using the tensor-perturbed metric
(\ref{tensormetric}), and then we Taylor expand the solution to order
$k^2$. Only the spatial part contributes to the tensor equation, and
by comparing it to the non-zero $k$ solution (\ref{nonzerok}), we
obtain two more constraints for the parameters $m_i$, given by
\be
m_1+2m_2+3m_5+4m_6=0, \qquad \qquad 9m_1+18m_2+24m_6=58.
\ee
These two additional equations allow us to fix the new constants,
\be
m_1=\frac{1}{2}-4m_0, \quad \quad m_2=\frac{1}{12}+2m_0 \quad\quad
m_5=-\frac{7}{3},\quad \quad m_6=\frac{13}{6},
\ee
with $m_0$ still undetermined, as before. 
One can extend this map to higher order $k$ by using the same
trick: the operator $(\nabla^\lambda e^{-\gamma})\nabla_\lambda
e^{m\gamma}$ applied on the previous $k$-order term, with the correct
$m$-number for a given $k$. To conclude this section, we have shown
that the first order $\alpha'$-corrections, in the ten-dimensional
string frame, are all accounted for by a non-trivial mapping from
eleven-dimensional Einstein gravity.

\section{Divergence of the $\alpha'$ series near the singularity}

In this final section, we want to understand in detail how the
$\alpha'$ expansion fails near the singularity. Because the
$(\nabla \phi)^8$ term at order $\alpha'^3$ has not, to our
knowledge, yet been computed, we shall only consider the effect
of the order $\alpha'$ term in the effective action. 
However, even with only this term, there are some interesting
features. Using the fact that
the dilaton and the scale factor
are related by
$a^2=e^{2\phi/3}$, the equations of motion for the background
(\ref{vara})-(\ref{varc}) can be reduced to 
to a single equation
\be\label{scalefactoreqn}
3\alpha'\big[(a')^4 -a\,(a')^2\,a''\big] +
a^4\big[(a')^2+a\,a''\big]=0,
\ee
where $'=\frac{d}{dt}$. This equation may be simplified by expressing
it in terms of the physical Hubble parameter $H = a'/a^2 $ and the
proper time $T= \int a(t) dt$; it becomes
\be\label{hubbleqn}
\dot{H}(1-3\,\alpha'\,H^2)+3\,H^2\,(1-\alpha'\,H^2)=0,
\ee
which is easily solved. For small $H$, {\it i.e.} in the regime
where the $\alpha'$ expansion is good, we have
$H\sim 1/3T$, implying $a\sim T^{1/3}\sim t^{1/2}$, as expected. For
large $H$, the solution goes like $H\sim 1/T$,
implying $a\sim T\sim e^{t}$. However, the equation fails at 
$T \sim 1$, because $dH/dT$ becomes infinite, so one
cannot get to the large $H$ regime. 

Nevertheless, we can certainly use equation (\ref{hubbleqn}) to compute
$H$ as a series in $\alpha'$; we obtain
\be\label{Hsol1}
H=\frac{1}{3T}+\frac{2\,\alpha'}{27\,T^3}+\frac{26\,\alpha'^2}{729\,T^5}
+\frac{242\,\alpha'^3}{10935\,T^7},
\ee
which one may check is equivalent to our earlier expresssions for
$a(t)$ in conformal time.

\begin{figure}[t!]
{\centering
\resizebox*{4in}{4in}{\includegraphics{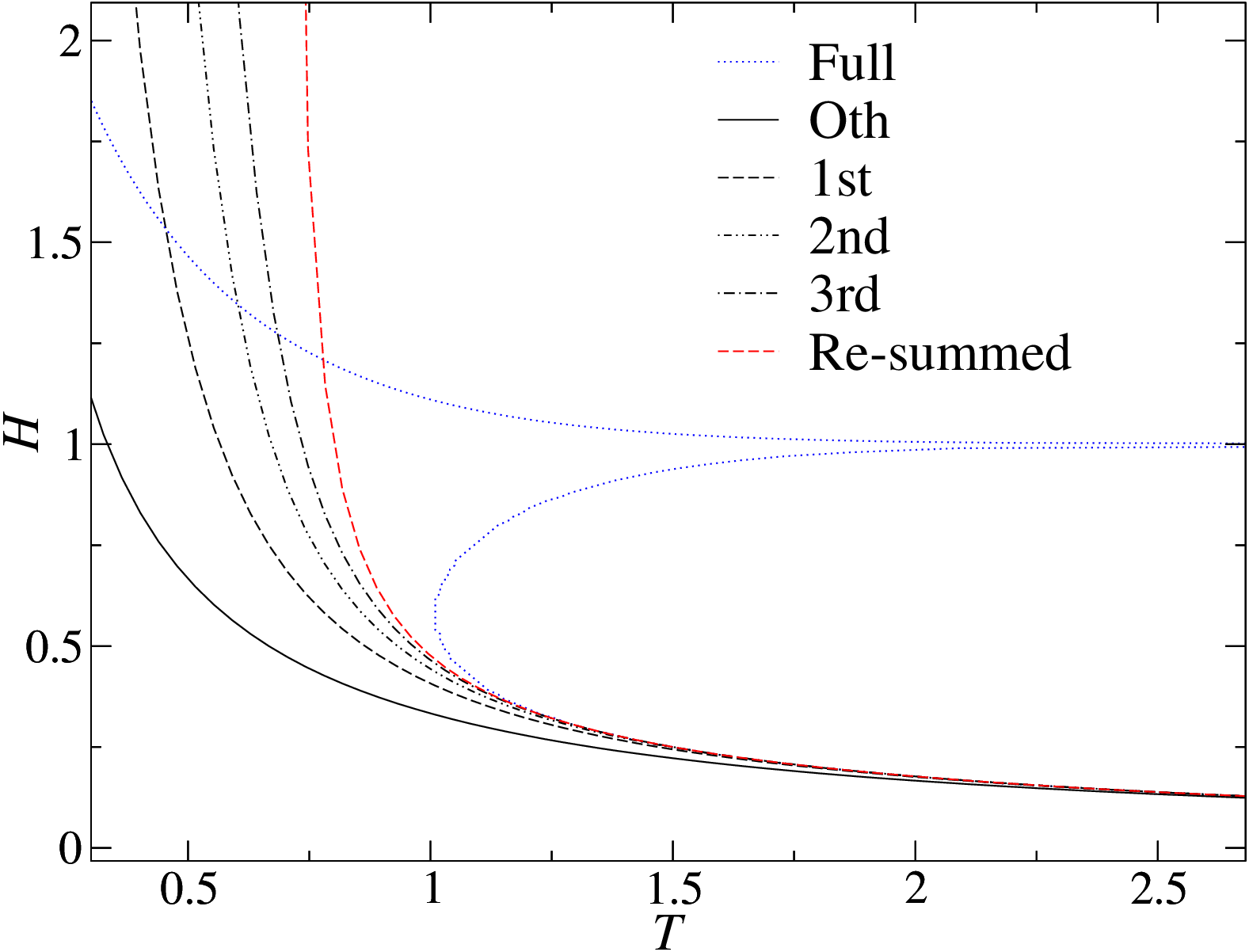}} }
\caption{Solutions to the Hubble equation (\ref{hubbleqn}). The series
  solution (\ref{Hsol1}) is presented order by order, and also its
  resummed expression using the Shank's transformation.
  The full solution to the Hubble equation
  (\ref{hubbleqn}) is also shown, being badly behaved when
  $\dot{H}=\infty$.}
\label{hubbleplot}
\end{figure}

Figure \ref{hubbleplot} shows the series solution (\ref{Hsol1}), order by
order in $\alpha'$, plotted against the solution of 
equation (\ref{hubbleqn}). Serious discrepancies set in around
$T\sim 1$, which of course comes as no surprise. 

Clearly, one cannot trust the calculation of the Hubble constant 
$H$ for times $T$ smaller than unity, in string units. If one
only has the series (\ref{Hsol1}), how would one go about 
checking this? One way is to attempt to resum the series;
we have chosen the Shanks 
transformation \cite{bender}, which is simple to apply in this case.
If a series has transient such that (in some region of the
complex plane) the
partial series $I_n$ is equal to $I+\lambda q^{n}$, where $\lambda
q^{n}$ is the transient and $|q|<1$, and $I_n\rightarrow I$ as  
$n\rightarrow\infty$, then we can use $I_{n-1}$, $I_n$ and
$I_{n+1}$ to calculate $I$, which is given by
\be\label{shanks}
I=\frac{I_{n+1}I_{n-1}-I_n^2}{I_{n+1}+I_{n-1}-2\,I_n}.
\ee 
Note, for example that this resummation is exact for the 
series $1+x+x^2+\dots= 1/(1-x)$. Applying this to the
series (\ref{Hsol1}), one finds a simple pole at $T \sim 0.7$. 
More important, however, the resummed curve is not very close to the 
series up to $\alpha'^3$ (see Figure \ref{hubbleplot}). Hence 
the resummation is unlikely to be correct. 

\begin{figure}[t!]
{\centering
\resizebox*{3in}{3in}{\includegraphics{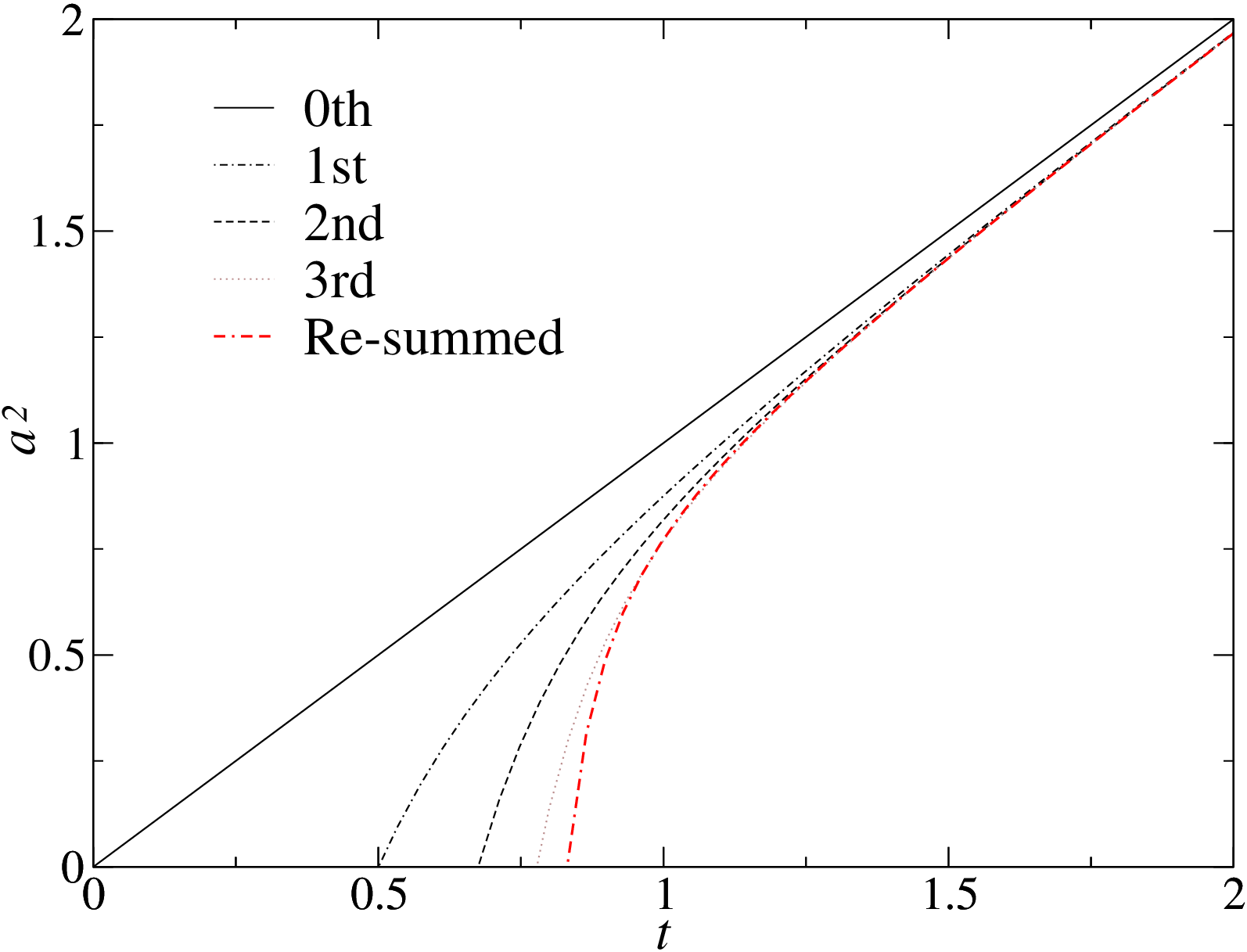}} 
\resizebox*{3in}{3in}{\includegraphics{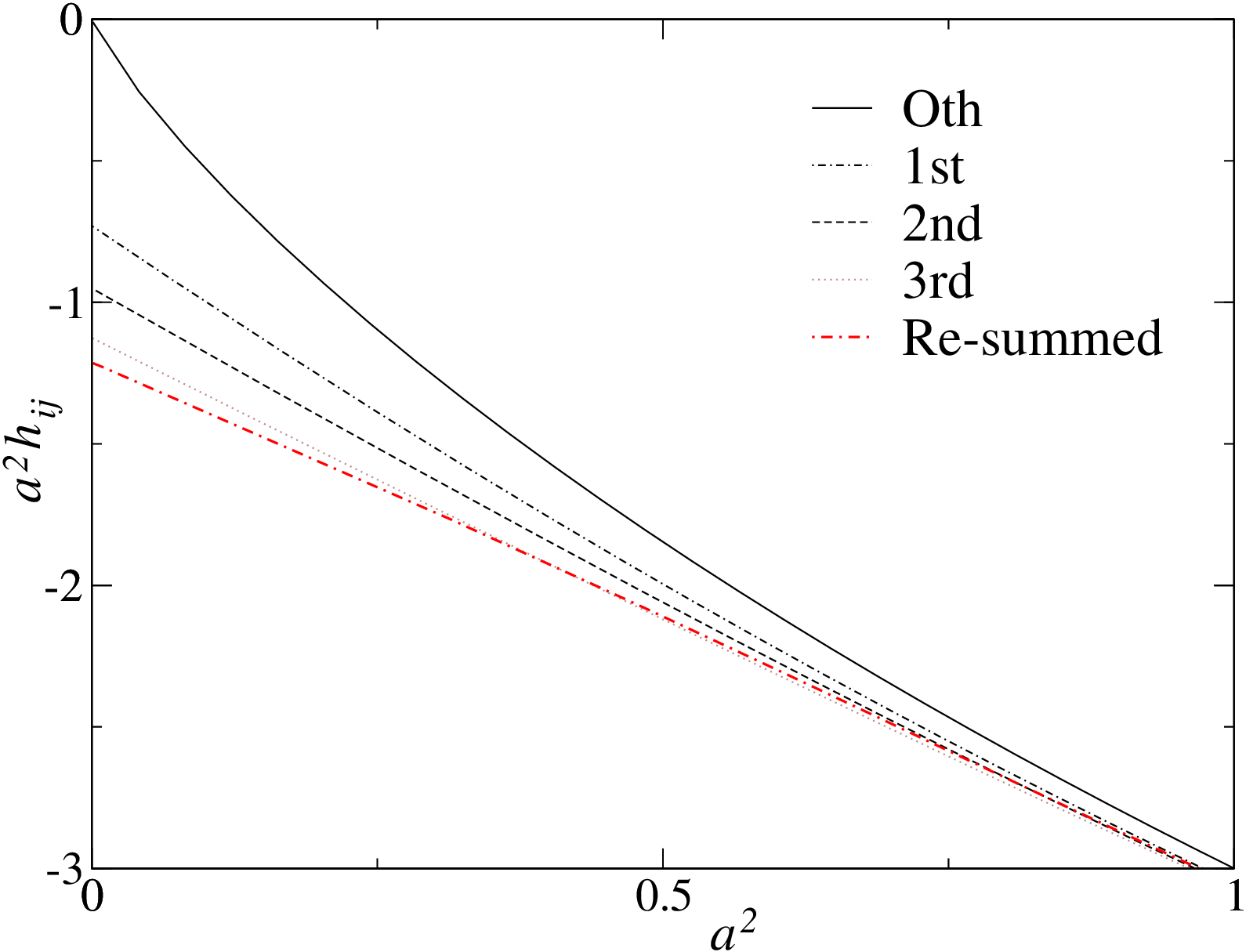}}}
\caption{The left shows how the scale factor reaches zero sooner when
  the $\alpha'$-corrections are included. In the right plot, one can
  see how the metric perturbation $a^2 h_{ij}$ remains finite all the
  way to singularity (\ie when $a\rightarrow 0$). The resummed
  expressions for both plots are also shown.} 
\label{h_vs_scale}
\end{figure}

However, some other quantities may be better behaved under
resummation. Let us return to consider the background
scale factor $a(t)$ and the 
 metric perturbations. As we have seen, the
singularity is reached earlier if the $\alpha'$-corrections are
included (see Figure \ref{h_vs_scale}). We can extrapolate the
metric perturbation $a^2 h_{\mu\nu}$ 
to this point in time to check whether
it is finite. To lowest order in $\alpha'$, it is zero, since 
$t \ln t$ tends to zero. This means one cannot match the 
amplitude of this term across $t=0$. However, as shown in 
Figure
\ref{h_vs_scale}, this quantity seems to remain finite 
as the singularity is approached. We can check these conclusions
by applying the same simple Shanks resummation, finding in this
case that the resummed value is rather close to the series
up to $\alpha'^3$. This is no more than suggestive, but
it may indicate that when all orders in $\alpha'$ 
are included these properties will persist.

\section{Conclusions}

In this paper, we have performed a detailed study of
$\alpha'$-corrections on the simplest ten-dimensional cosmological 
background solution in string theory, corresponding to eleven-dimensional 
compactified Milne spacetime in M-theory. We have computed 
the effect of these corrections on both the background and on 
linearized scalar and tensor perturbations, to first order in
$\alpha'$. From the M-theory viewpoint, away from the singularity, our
background is Riemann-flat in eleven dimensions and hence it should
provide an exact  
background to M-theory. Similarly, one can argue that the linearized
perturbations should be exactly described by Einstein gravity in
eleven dimensions, since any higher powers of curvature invariants
would give no contribution. As a check of this idea, we have verified
that, to the order in $\alpha'$ we compute, all  
the $\alpha'$-corrections in the string frame calculations are indeed
possible to generate by field redefinitions. Several ``miracles" are
necessary in order for this to occur - for example, the dilaton and
the scale factor get corrections in just such a way that the right
combination, representing the ``transverse" scale factor in eleven
dimensions, remains unchanged.  

From the string theory viewpoint, we have shown that the
$\alpha'$-corrections modify the usual zeroth order solutions so that
the singularity occurs sooner, in background conformal time, for both
the metric and dilaton fields. Thus the $\alpha'$-corrections do not
seem to resist the formation of a cosmic singularity. We have
attempted a simple resummation, obtaining a simple pole as the
asymptotic behaviour near $t=0$, and a finite metric perturbation when
the scale factor reaches zero. However, such results are inconclusive,
because we expect $\alpha'$ expansion to break down well before the
singularity, which may be describeable using an expansion in
$1/\alpha'$, as discussed in \cite{turok, niz}.  

Clearly, we cannot use the $\alpha'$ expansion to study the singularity
itself, nor the transition across it. It will be necessary to develop a
more powerful calculational approach, incorporating both the $\alpha'$ 
expansion at large times, and a new expansion in $1/\alpha'$ near the
singularity. Nevertheless, it may be hoped that the results we have
obtained will be useful in comparing with this future, more complete
treatment.

\subsection*{Acknowledgements}
We thank Michael Green, Hugh Osborn, Malcolm Perry, Aninda Sinha and
Paul Townsend for useful discussions and pointing out several relevant 
references. This work was supported by CONACYT, SEP and St Edmund's
College (GN) and PPARC (NT).

\end{document}